\documentclass[aps,reprint]{revtex4-1}
\pdfoutput=1

\usepackage{braket}
\usepackage{amsmath}
\usepackage{amsthm}
\usepackage{amssymb}
\usepackage{mathrsfs}
\usepackage{mathtools}
\usepackage{color}
\usepackage{comment}

\makeatletter

\newcommand{\Rmnum}[1]{\expandafter\@slowromancap\romannumeral #1@}
\makeatother

\begin{document}

\title{Lamb shift multipolar analysis}

\author{Emmanuel Lassalle}
\affiliation{Aix Marseille Univ, CNRS, Centrale Marseille, Institut Fresnel, Marseille, France}

\author{Alexis Devilez}
\affiliation{Aix Marseille Univ, CNRS, Centrale Marseille, Institut Fresnel, Marseille, France}

\author{Nicolas Bonod}
\affiliation{Aix Marseille Univ, CNRS, Centrale Marseille, Institut Fresnel, Marseille, France}

\author{Thomas Durt}
\affiliation{Aix Marseille Univ, CNRS, Centrale Marseille, Institut
  Fresnel, Marseille, France}

\author{Brian Stout}
\email[]{brian.stout@fresnel.fr}
\affiliation{Aix Marseille Univ, CNRS, Centrale Marseille, Institut Fresnel, Marseille, France}

\date{\today}

\begin{abstract}
It is now well established that radiative decay of quantum emitters can be strongly modified by their environment.
In this paper we present an exact --- within the weak-coupling approximation
--- multipole expression to compute
the Lamb (frequency) shift induced by an arbitrary set of resonant scatterers on a
nearby quantum emitter, using multi-scattering theory. We
also adopt a Quasi-Normal Mode description to account for the line shape of the
Lamb shift spectrum in the
near-field of a plasmonic nanosphere. It is then shown
that the Lamb shift resonance can be blue-shifted
as the size of the nanoparticle increases, suggesting that
nanoparticles may be used to tune this resonant interaction. Finally,
a realistic calculation of the Lamb shift is made for a dimer configuration. 
\end{abstract}

\maketitle

\section{Introduction}
Control of the decay properties of quantum emitters \emph{via} modifications
of their local electromagnetic environment is
being actively pursued due to the rich perspectives it offers for both
fundamental and practical applications \cite{tame2013quantum}.
In the weak-coupling regime, the exponential decay in time of the
excited state is characterized by the decay rate, for which it is well
known that it can be either enhanced \cite{purcell1946spontaneous} or
inhibited \cite{hulet1985inhibited} by the local electromagnetic (EM)
environment. With the convergence of communities such as
near-field optical microscopy, semiconductors, plasmonics, and metamaterials, engineering the quantum vacuum allows
tailoring the decay rate in unprecedented ways \cite{PhysRevLett.115.025501, PhysRevApplied.6.064016}.
A less often discussed effect of spontaneous emission is that the
surrounding environment also induces level shifts of the excited atomic states, resulting in a
frequency-shift for the emitted photons, in comparison with the bare resonance frequency. This is the so-called \emph{Lamb shift},
which originally refered to level shifts of atoms in free
  space \cite{PhysRev.72.241, bethe1947electromagnetic}, also called
  radiative frequency-shift or Casimir-Polder frequency-shift. This
  effect has been theoretically studied in the case of perfect reflectors
\cite{barton1974frequency}, partially reflecting surfaces
\cite{PhysRevA.12.1448,PhysRevA.32.2030} and photonic crystals \cite{kofman1994spontaneous,PhysRevLett.84.2136,PhysRevLett.93.073901}. 
Multipole formulas of the
Lamb shift have been
derived in the case of a dielectric microsphere without 
\cite{ching1987dielectric,klimov1996radiative} and with
\cite{PhysRevA.64.013804} absorption, and for dielectric or metallic prolate spheroids \cite{klimov2002spontaneous}. However, there
is no such formulas in multi-scattering configurations, except in the case of two-dimensional photonic
crystals~\cite{asatryan2006frequency}.

In this article, we derive --- using the generalized Mie theory
\cite{doi:10.1080/09500340210124450, stout2008recursive} --- a multipole
formula for the Lamb shift of a quantum emitter induced by an
arbitrary set of scatterers.
This formula is exact within the weak-coupling approximation and does not
take into account non-local effects which come into play for emitter - particle distances below one nanometer \cite{zhu2016quantum}.

Section \ref{sec:classic_vs_quantum} justifies the use of a classical
formalism to study the Lamb shift induced by the presence of matter by showing, in 
the weak-coupling approximation, its equivalence to the fully quantum result.
An exact multipole formula for the Lamb shift is then derived in section
\ref{sec:T-Matrix} and illustrated in section \ref{sec:analysis} by computing the Lamb
shift in the vicinity of a silver nanosphere, where we also show that
the spectral line shape of the Lamb shift can be accounted for in the context of a ``Quasi-Normal Mode''
description.
In section \ref{predictions}, we study the influence of the
nanoparticle's size on the environmentally induced Lamb shift, and we
predict a displacement of the emitter's 
Lamb shift resonance as the size of the nanoparticle
changes. Finally, as a practical calculation, we compute the Lamb shift in the case of a dimer nanoantenna.

\section{Environmentally induced Lamb shift}
\label{sec:classic_vs_quantum}

\subsection{Classical approach}

An excited two-level atom with transition frequency
$\omega_0$ and natural linewidth $\gamma_0$ can be modeled by a harmonically
oscillating \emph{point dipole}, whose electric dipole moment
$\bold{p}(\bold{r}_0,t)$ obeys, in the case of
\emph{small damping} ($\gamma_0 \ll \omega_0$) \cite{novotny2012principles}:
\begin{equation}
\frac{d^2\bold{p}(\bold{r}_{0},t)}{dt^2} + \gamma_{0}\frac{d\bold{p}(\bold{r}_0,t)}{dt} +
\omega_0^2\bold{p}(\bold{r}_0,t) = \frac{q^2}{m}\bold{E}_s(\bold{r}_0,t) \;,
\end{equation}
where $\{\omega_0$, $\gamma_0$, $q$, $m\}$ are the characteristics of the
classical dipole (the natural frequency of the oscillator, the damping constant
in the \emph{homogeneous} background, the charge and the mass respectively) and
$\bold{E}_s(\bold{r}_0,t)$ is the field scattered by the environment
at the dipole position $\bold{r}_0$.
Adopting the following \emph{ansatz}:
\begin{equation}
\small \left\{
    \begin{array}{ll} 
      \bold{p}(\bold{r},t) = \bold{p}_0e^{-\mathrm{i}\Omega t}\\
      \bold{E}_s(\bold{r}_0,t) = \bold{E}_s(\bold{r}_0,\omega_0)e^{-\mathrm{i}\Omega t}
\end{array}
\right. \text{where} \left\{
    \begin{array}{ll} 
      \Omega = \omega_0 + \Delta\omega - \mathrm{i}\frac{\gamma}{2}\\
      \Delta \omega = \omega - \omega_0 
\end{array} 
\right. \;,
\end{equation}
with $\gamma$ and $\omega$ respectively indicating the new decay rate and resonance
frequency, together with the \emph{weak-coupling approximation} in a
classical context,
\begin{equation}
\frac{q^2}{m}|\bold{E}_s| \ll \omega_0^2 |\bold{p}| \ ,
\label{eq:classical_wc}
\end{equation}
one finds the following expression for the frequency-shift of the light
emitted by the dipole due to the environment \cite{novotny2012principles}:
\begin{equation}
\left . \frac{\Delta\omega}{\gamma_0} \right |_{\omega_0} =
-\frac{3\pi\epsilon_0\varepsilon_b}{k^3}\times\frac{1}{|\bold{p}_0|^2}\times
\text{Re}(\bold{p}_0^*\cdot\bold{E}_s(\bold{r}_0,\omega_0)) \;,
\label{eq:classical_freq_shift}
\end{equation}
where $k=n_b(\omega_0/c)$ is the wave-number of the nonabsorbing homogeneous 
background medium of refractive index $n_b=\sqrt{\varepsilon_b}$. In
this classical picture, one
can see from Eq.~(\ref{eq:classical_freq_shift}) that the environment
contribution to the frequency-shift is due to the dipole
interacting with its own electric field scattered back
by the environment.

To link this expression with the quantum one, one can derive the dipole fields using the
\emph{Green-function} formalism (for the sake of simplicity, we
consider the dipole emitter to be in vacuum: $\varepsilon_b = 1$). The
field produced at $\bold{r}$ by a \emph{point} dipole
located at $\bold{r}_0$ and with natural frequency $\omega_0$ is \cite{novotny2012principles}:

\begin{equation}
\bold{E}(\bold{r},\omega_0)
=\omega_0^2\mu_0\overset\leftrightarrow{\bold{G}}(\bold{r},\bold{r}_0,\omega_0)\cdot \bold{p}_0\;,
\label{eq:green_function}
\end{equation}
where $\overset\leftrightarrow{\bold{G}}$ denotes the dyadic Green tensor.
\noindent
By separating the Green tensor into an ``unperturbed'' $\overset\leftrightarrow{\bold{G}}_0$ plus a
``scattering'' $\overset\leftrightarrow{\bold{G}}_s$ contributions
\cite{novotny2012principles},
\begin{equation}
\overset\leftrightarrow{\bold{G}} =
\overset\leftrightarrow{\bold{G}}_0 +
\overset\leftrightarrow{\bold{G}}_s \;,
\label{eq:GF_contrib}
\end{equation}
Eq.~(\ref{eq:classical_freq_shift}) can be cast in terms of the
scattering Green tensor:
\begin{equation}
\left. \frac{\Delta\omega}{\gamma_0} \right |_{\omega_0} = - \frac{3\pi
  c}{\omega_0}\times \bold{u}_p\cdot
\text{Re}(\overset\leftrightarrow{\bold{G}}_s(\bold{r}_0,\bold{r}_0,\omega_0))\cdot\bold{u}_p \;,
\label{eq:classic}
\end{equation}
with $\bold{u}_p$ being the unit vector in the direction of the
dipole moment: $\bold{p}_0=p_0\bold{u}_p$.

\subsection{Quantum approach}

In a quantum approach, the excited two-level atom is modeled by its
state vector $\ket{e}$, and its interaction with the electromagnetic field is represented by an
interaction Hamiltonian $\hat{H}_I$. The \emph{weak-coupling approximation}
in a quantum context consists of considering that the matrix elements of the interaction Hamiltonian
are small compared to those of the non-interacting Hamiltonian $\hat{H}_0$.
Therefore, the energy level shift $\Delta E$ of the excited
atomic state is calculated by using the usual perturbation theory to second order in
the perturbation $\hat{H_I}$. Besides, by using the fluctuation-dissipation
theorem, one can show that the energy-shift of the first excited state $\ket{e}$ of bare frequency $\omega_0$ is \cite{PhysRevA.32.2030}:
\begin{equation}
\left. \Delta E \right |_{\omega_0} = - \frac{\omega^2_0}{\pi\epsilon_0c^2} \, p_ip_j \, \mathcal{P} \left [ \int_{0}^{+\infty} \mathrm{d}\omega \,
\frac{\text{Im}\left(G_{ij}(\bold{r}_0,\bold{r}_0,\omega)\right)}{\omega
  - \omega_0}\right ] \;,
\label{eq:deltaE}
\end{equation}
where $\mathcal{P}$ denotes the principal value of the integral,
$\bold{p} = \bra{g}\hat{\bold{p}}\ket{e}$ ($\hat{\bold{p}}$ being the dipole
moment operator and $\ket{g}$ the ground state vector) is the transition dipole matrix
element, and $G_{ij}$ is the previous classical Green
tensor (let us note that the notation $\overset\leftrightarrow{\bold{G}}$ used
in \cite{PhysRevA.32.2030} is the field susceptibility that we call
$\overset\leftrightarrow{\bold{F}}$, and which is related to the Green
tensor by $\overset\leftrightarrow{\bold{F}}(\bold{r},\bold{r}',\omega) \leftrightarrow
\omega^2\mu_0\overset\leftrightarrow{\bold{G}}(\bold{r},\bold{r}',\omega)$).
By using the Kramers-Kronig relations for the Green tensor, and
separating as previously the Green tensor into two contributions, one can cast the frequency-shift resulting from the
energy level shift induced by the presence of matter, in the form:
\begin{equation}
\left. \Delta \omega \right |_{\omega_0}= - \frac{\omega^2_0}{\hbar\epsilon_0c^2}\,p_ip_{j}\,\text{Re}\left((G_s)_{ij}(\bold{r}_0,\bold{r}_0,\omega_0)\right)
+ \text{QC} \;.
\label{eq:quantum}
\end{equation}
Except for the non-resonant quantum correction term QC which is negligibly
small \cite{PhysRevA.64.013804}, this expression has the same form as the
classical formula provided that one normalizes by the quantum
decay rate in free space
\begin{equation}
\gamma_0 = \frac{\omega^3_0|\bold{p}|^2}{3\pi\epsilon_0\hbar c^3} \;,
\end{equation}
because the
normalization eliminates the dependency on $\bold{p}$
and provides a safe link between quantum and classical
formalisms.

Thus, in the \emph{weak-coupling regime}, the quantum treatment gives the same result
as the classical treatment when considering the normalized
frequency-shift --- that we will call Lamb shift in the following --- \emph{between the ground state and the first excited state} (to
consider other atomic levels, the classical
treatment and the two-level atom model fail, and one must refer to the
general formula derived in \cite{PhysRevA.32.2030}).
Note that for an \emph{absorbing} medium, characterized by an imaginary part of
its permittivity, this equivalence still holds, because on one hand, in the classical approach developed in terms of the
Green tensor, the permittivity can become complex, and in a quantum context, the link between the ground-state fluctuations of the
electric field and the classical Green tensor remains the same \cite{PhysRevA.57.3931,PhysRevA.58.700,PhysRevA.64.013804}.

\section{Multipole formula for the Lamb shift}
\label{sec:T-Matrix}
Now we move to the derivation of the exact multipole formula for the
Lamb shift induced by an arbitrary set of resonant scatterers on a nearby quantum emitter.
One can see from Eq.~(\ref{eq:classical_freq_shift}) that the Lamb
shift induced by the surrounding environment is embodied in the field
scattered by the environment $\bold{E}_s$, which can be calculated
from the scattering part $\overset\leftrightarrow{\bold{G}}_s$ of the
total Green tensor through Eqs.~(\ref{eq:green_function}) and (\ref{eq:GF_contrib}).
The determination of $\overset\leftrightarrow{\bold{G}}_s$ is thus the chief obstacle to the calculation
of the Lamb shift.
From a classical viewpoint, the scattering Green tensor $\overset\leftrightarrow{\bold{G}}_s$ must
take into account the multiple scattering of the incident radiation
from all the scatterers. Therefore, for the purpose
of calculation, it is advantageous to express the scattering Green
tensor in terms of the multiple-scattering T-Matrix \cite{stout2011multipole}, where the T-Matrix is defined in operator notation as 
\begin{equation}
\overset\leftrightarrow{\bold{G}}_s =
\overset\leftrightarrow{\bold{G}}_0 \,
\overset\leftrightarrow{\bold{T}} \,
\overset\leftrightarrow{\bold{G}}_0 = \overset\leftrightarrow{\bold{G}}_0 \,
\left(\sum_{i=1, j=1}^N \overset\leftrightarrow{\bold{T}}^{(i,j)}
\right) \, \overset\leftrightarrow{\bold{G}}_0 \;,
\label{eq:GTG}
\end{equation}
and has been split into $N^2$ operators
$\overset\leftrightarrow{\bold{T}}^{(i,j)}$ (that represent all
multiple-scattering events from a
multiple-scattering viewpoint \cite{doi:10.1080/09500340210124450}), $i$ and $j$ being the particle labels, and $N$ the total number of
scatterers. 

In order to calculate the $\overset\leftrightarrow{\bold{T}}^{(i,j)}$ operators, we will make
use of the multipolar fields --- also called multipolar modes or
multipoles --- which are a set of basis EM modes that are especially useful in describing EM
scattering for particles with spherical symetries \cite{zambrana2015control}. We will denote a
multipolar field as $\ket{\bold{\Psi}_{q,n,m}}$, each mode being specified by three
discrete numbers: $q$ accounts for the parity of the field, and $q=1$ for
a magnetic mode and $q=2$ for an electric mode; $n=1,2,...,\infty$ and
will be called the ''multipolar
order''; and $m=-n,...,n$ and will be called the ''orbital number''. Explicit representations of these modes can be found in
\cite{zambrana2015control}, and here the fields and
operators will be expressed in the basis of the multipolar fields satisfying the outgoing boundary conditions (called the Hankel multipolar
fields in \cite{zambrana2015control}), that we will note
$\bold{M}_{nm}(k\bold{r})$ for the magnetic modes ($q=1$) and $\bold{N}_{nm}(k\bold{r})$ 
for the electric modes ($q=2$) in the real space
representation.

The $\overset\leftrightarrow{\bold{T}}^{(i,j)}$ operators are
then expressed in the multipole basis~\cite{stout2011multipole}:
\begin{equation}
\tiny
\overset\leftrightarrow{\bold{T}}^{(i,j)} =
\sum_{q,q'=1}^2\sum_{n,n'=1}^\infty\sum_{m=-n}^n\sum_{m'=-n'}^{n'}\ket{\bold{\Psi}_{q,n,m}}T_{q,n,m;q',n',m'}^{(i,j)}
\bra{\bold{\Psi}_{q',n',m'}} \;,
\end{equation}
and can be calculated from the infinite dimensional $T^{(i,j)}$ matrices, that can be rendered
finite by truncating the multipolar order $n$ to some finite dimension
$n_{\text{cut}}$ (the choice of $n_{\text{cut}}$ for which the summation with
respect to the multipolar order $n$ converges will depend on particle size and interaction
strengths). Several methods exist for calculating the $T^{(i,j)}$
matrices, and we use the analytical balancing techniques
detailed in \cite{stout2008recursive} and implemented in an in-house
code used for the numerical simulations of this article.
Once the on-shell $T^{(i,j)}$ matrices have been determined, one can
compute the expression of the electric field $\bold{E}_s(\bold{r}_0,\omega_0)$ scattered by the environment
by employing Eq.~(\ref{eq:GTG}) in Eqs.~(\ref{eq:GF_contrib}) and (\ref{eq:green_function}):
\begin{equation}
\small
\bold{E}_s(\bold{r}_0,\omega_0)  = \frac{\mathrm{i}p_0k\omega_0^2}{\epsilon_0c^2}\sum_{i,j=1}^N
\left[[\bold{M}(k\bold{r}_i),\bold{N}(k\bold{r}_i)]^t T^{(i,j)}H^{(j,0)} f\right] \;,
\end{equation}
where $[\bold{M} , \bold{N}]$ is a column matrix composed of the
$\bold{M}_{nm}$ and $\bold{N}_{nm}$ functions, $f$ represents the dipolar source and denotes a column
matrix containing the emitter coefficients in
the multipole space, and $H^{(j,0)}$ is the
irregular translation-addition matrix between the emitter position at $\bold{r}_0$
and the position of particle $j$ (for more details, see the derivation
of Eq.~(19) in \cite{stout2011multipole}). 

Finally, the expression of $\bold{E}_s(\bold{r},\omega_0)$ can be utilized in
Eq.~(\ref{eq:classical_freq_shift}) to obtain the multipole expression
for the normalized Lamb shift induced by the presence of $N$ scatterers:
\begin{equation}
\frac{\Delta\omega}{\gamma_0} =
3\pi\times \,\text{Im}\left(\sum_{i,j=1}^N f^t
  H^{(0,i)}T^{(i,j)}H^{(j,0)} f\right) \;.
\label{eq:num_sim}
\end{equation}
In the case of a single particle ($N=1$), Eq.~(\ref{eq:num_sim})
takes the form:
\begin{equation}
\frac{\Delta\omega}{\gamma_0} =
3\pi\times \text{Im}\left(f^t H^{(0,1)}tH^{(1,0)} f\right) \;,
\label{eq:Mie}
\end{equation}
where $t$ is the single-particle T-Matrix. In the case
of a spherical Mie scatterer, $t$ is a diagonal matrix composed of the Mie
coefficients of the sphere (given in Appendix~\ref{App:A}), and Eq.~(\ref{eq:Mie}) is then equivalent to expressions previously derived
for a single sphere \cite{klimov1996radiative,PhysRevA.64.013804}. Exact analytical
expressions of the first two multipolar contributions to the
Lamb shift can be
found in Appendix~\ref{App:B}.

\section{Multipolar analysis}
\label{sec:analysis}

\subsection{Multipole contributions to the Lamb shift}
\begin{figure}
\centering
\includegraphics[width=\linewidth]{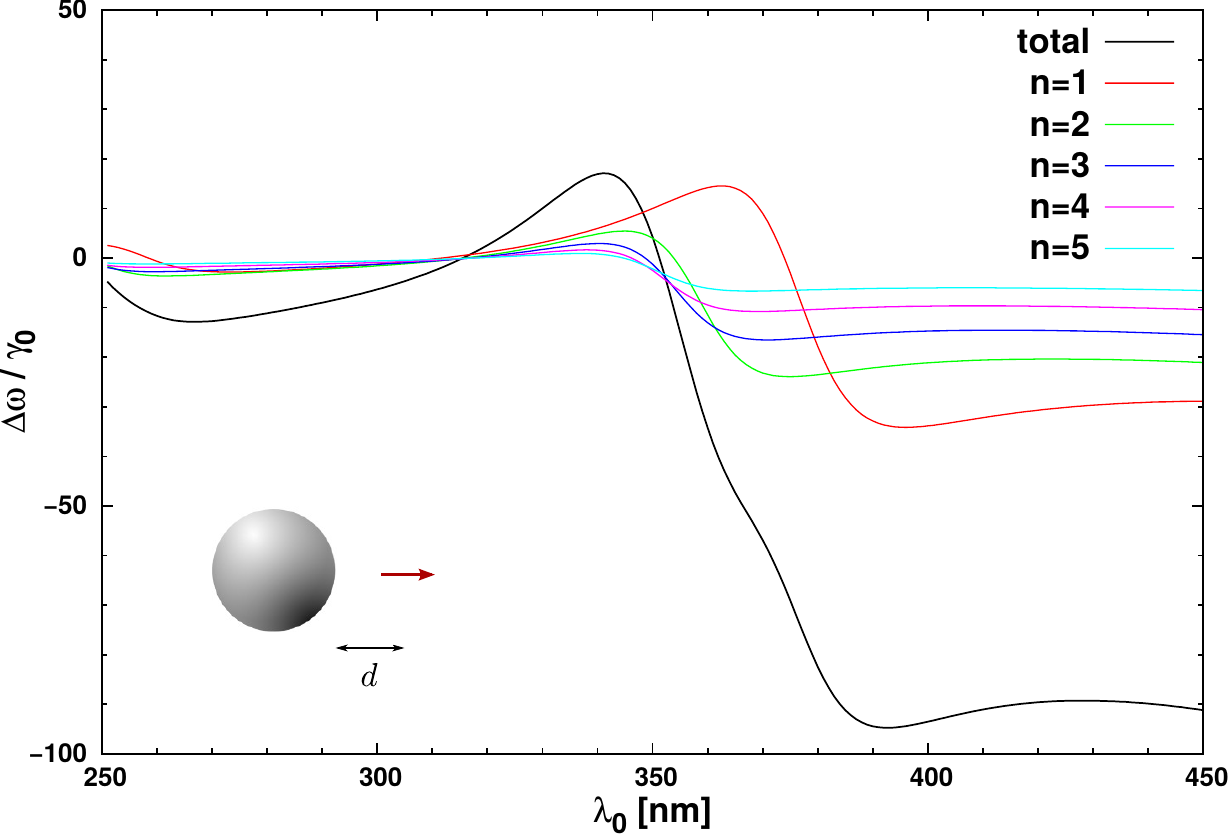}
\caption{Numerical simulations of the total Lamb shift $\Delta
  \omega$ (black curve) and its multipolar
  contributions $n = 1,2,3,4,5$ (colored curves) as a function of the transition
  wavelength $\lambda_0 = 2\pi c/\omega_0$ for a
  perfect electric dipole emitter with radial orientation and located at
  $d = 10\,\text{nm}$ from a silver nanosphere with $a = 20\,\text{nm}$
  radius (red arrow). The Lamb shift is normalized to the dipole's decay rate in
  free space $\gamma_0$. The refractive index of the
homogeneous background is $n_b=1$. A Drude-Lorentz
model for the silver permittivity is used according to
\cite{rakic1998optical}. The total Lamb
shift is computed by taking $n_{\text{cut}}=10$.}
\label{fig:multipolar_contribution}
\end{figure}

Let us first calculate the Lamb shift in the case of a silver
nanosphere of radius $a=20\,\text{nm}$ in vacuum ($n_b=1$). Based on Eq.~(\ref{eq:Mie}), we compute using an in-house code the
Lamb shift of a quantum emitter radially oriented and located at a
distance $d=10\,\text{nm}$ from the nanoparticle, as a function of the bare transition wavelength
$\lambda_0 = 2\pi c/\omega_0$ (black curve in
Fig.~\ref{fig:multipolar_contribution}). We analyze this Lamb shift
spectrum by plotting separately the
different multipolar contributions
(plotted in colors in Fig.~\ref{fig:multipolar_contribution}: $n=1$
corresponds to the contribution of the dipolar mode, $n=2$ to the
contribution of the quadrupolar
mode and so on). One can thus
see that in the near-field of the
nanoparticle, the total Lamb shift is due to the contribution of
several multipolar modes and the fact that the dipole approximation to model the response of the
nanoparticle (corresponding to the red curve in Fig.~\ref{fig:multipolar_contribution}) fails to account for
the Lamb shift.
In other words, in the near-field region, the atom couples to several plasmon
modes of the silver nanoparticle (see also \cite{ColasdesFrancs:08}), which gives rise to the complex
pattern of the Lamb shift spectrum. 

\begin{figure}
\centering
\includegraphics[width=\linewidth]{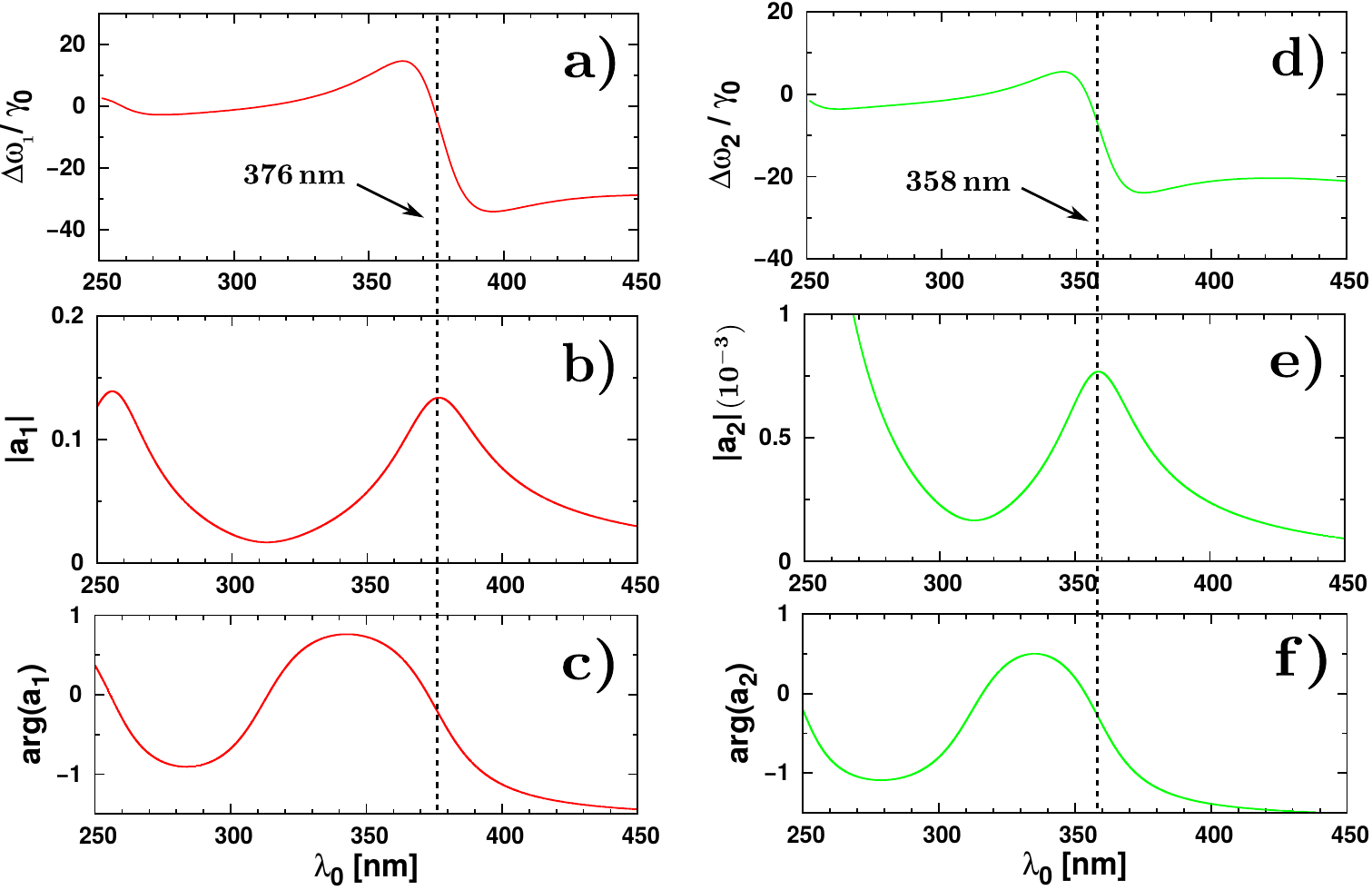}
\caption{(a) and (d): Lamb shift dipolar (red curve) and quadrupolar
  (green curve) contributions of Fig.\ref{fig:multipolar_contribution} (same color
  code) normalized by $\gamma_0$. (b) and (e): Modulus of the associated electric dipolar (red curve) and
  quadrupolar (green curve) Mie coefficients $a_1$ and $a_2$ as a function of the
  excitation wavelength $\lambda_0$. (c) and (f): Argument
  of the associated electric dipolar (red curve) and quadrupolar
  (green curve) Mie coefficients
  $a_1$ and $a_2$ as a function of $\lambda_0$. A Drude-Lorentz
model for the silver permittivity is used according to
\cite{rakic1998optical}.}
\label{fig:mie_coeff}
\end{figure}

In order to account for the spectral
line shape, we will make use of the analytical expressions of the
dipolar and quadrupolar contributions derived in Appendix
\ref{App:B}, in the case of a radially oriented dipole. In the non-retarded 
regime $k d\ll1$ (which is fulfilled here), the analytical
expression of the dipolar contribution ($n=1$) reduces to,
\begin{equation}
\frac{\Delta\omega_1^\perp}{\gamma_0} =
\frac{9}{2}\frac{1}{(kd)^6}\text{Im}[ a_1] + O\left( (kd)^{-6} \right) \;,
\end{equation}
while the quadrupolar contribution ($n=2$) reduces to,
\begin{equation}
\frac{\Delta\omega_2^\perp}{\gamma_0} =
\frac{405}{2}\frac{1}{(kd)^8}\text{Im}[a_2] + O\left( (kd)^{-8} \right) \;,
\end{equation}
where the subscript $\perp$ indicates a dipole perpendicular to the
particle surface (radially oriented), and $a_1$ ($a_2$) is the electric dipolar (quadrupolar) Mie
coefficient whose expression can be found in Appendix~\ref{App:A}.
The explanation of the spectral behavior of the Lamb shift
is thus found in the imaginary part of the Mie coefficient. In
Fig.~\ref{fig:mie_coeff}, we plot the modulus ((b) and (e)) and phase
((c) and (f)) of the
electric dipolar and quadrupolar Mie coefficients $a_1$ and $a_2$
respectively as a function of the excitation wavelength, together with the first two multipolar
contributions $n=1$ and $n=2$ of Fig.~\ref{fig:multipolar_contribution} ((a) and (d)
in Fig.~\ref{fig:mie_coeff} plotted with the
same color code). One can see that the inflection point of
the Lamb shift spectrum (around
$376\,\text{nm}$ for $n=1$ and $358\,\text{nm}$ for $n=2$) corresponds to a
resonance maximum of the modulus of the associated Mie coefficient accompanied by a strong
phase change (the resonance of the Mie coefficients around $250\,\text{nm}$ is a
  spurious resonance peculiar to the model of permittivity used \cite{hao2007efficient}). This clearly shows the multipolar origin of the plasmon
resonance enhanced Lamb shift. 


\subsection{Quasi-normal mode description}
Another interpretation of the shape of the Lamb shift spectrum can be
given using a Quasi-Normal Mode (QNM) description \cite{PhysRevLett.110.237401} 
(also called ``Resonant State'' expansions). By expanding the scattered field
$\bold{E}_s(\bold{r}_0,\omega_0)$ in Eq.~(\ref{eq:classical_freq_shift}) onto a
small set of QNMs of the plasmonic resonator as in \cite{yang2015analytical}, we obtain:
\begin{equation}
\left. \frac{\Delta\omega}{\gamma_0}\right |_{\omega_0} \simeq
\sum_{\alpha} A_\alpha\left(\frac{\omega_\alpha'}{\omega_0}\right)^2\frac{\omega_\alpha ''}{2}\frac{\omega_\alpha'-\omega_0}{(\omega_\alpha'-\omega_0)^2 + \omega_\alpha''^2} + B_\alpha(\omega_0) \;,
\label{eq:omega_nf}
\end{equation}
where $\omega_\alpha = \omega_\alpha' + \text{i} \,\omega_\alpha''$ is the complex
frequency of the QNM labeled $\alpha$, while $A_\alpha$ is a dimensionless
factor and $B_\alpha(\omega_0)$ a function of
$\omega_0$ (for the qualitative analysis which follows, we will consider it
as constant: $B_\alpha(\omega_0) \equiv B_\alpha$). 
An equivalent expression in term of the wavelength is obtained by
extending the relation between $\omega$ and $\lambda$ to complex
numbers. Adopting $\lambda_\alpha \equiv 2\pi c /
\omega_\alpha$, where $\lambda_\alpha = \lambda_\alpha' + \text{i}\,
\lambda_\alpha''$ is the complex wavelength
associated with the complex frequency, $\omega_\alpha = \omega_{\alpha}' +\text{i}\,\omega_{\alpha}''$, we find:
\begin{equation}
\left. \frac{\Delta\omega}{\gamma_0} \right |_{\lambda_0} \simeq \sum_{\alpha} -A_\alpha\left(\frac{\lambda_\alpha'}{\widetilde{\lambda}_0}\right)^2\frac{\lambda_\alpha''}{2}\frac{\lambda_\alpha'-\widetilde{\lambda}_0}{(\lambda_\alpha'-\widetilde{\lambda}_0)^2 + \lambda_\alpha''^2} + B_\alpha \;,
\label{eq:lambda_nf}
\end{equation}
where $\widetilde{\lambda}_0 \equiv
|\lambda_\alpha|^2/\lambda_0$. Note that
Eqs. (\ref{eq:omega_nf}) and (\ref{eq:lambda_nf}) are generally valid for any
resonator shape, and shows that the total Lamb shift can be given by the
sum of independent contributions of the QNMs.

\begin{figure*}[htbp]
\centering
\includegraphics[width=\linewidth]{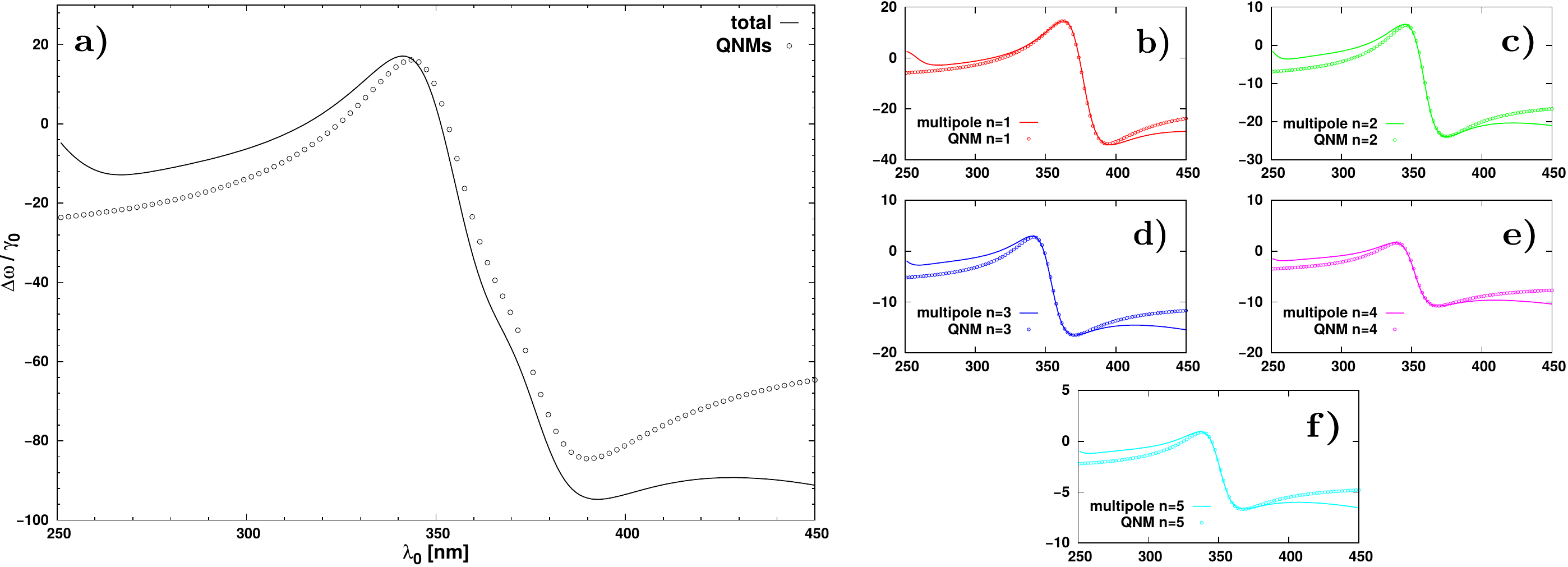}
\caption{Comparison between numerical simulations
  and analytical calculations of the Lamb shift. (a): Total Lamb shift of
Fig.~\ref{fig:multipolar_contribution} (black curve) compared to the Lamb
  shift calculated with Eq.~(\ref{eq:lambda_nf_bis}) using the five QNM
  resonances displayed in Table~\ref{tab:coeff} (dotted curve). (b) to
  (f): Fit of each multipole contribution of
  Fig.~\ref{fig:multipolar_contribution} (full lines, same color code) with the
  corresponding QNM contribution calculated with Eq.~(\ref{eq:lambda_nf_bis})
  (dotted lines) with $A_n$ and $B_n$ the fitting parameters (displayed in Table~\ref{tab:coeff}).}
\label{fig:QNM_contribution}
\end{figure*}

For a spherical Mie resonator, the QNMs are the
multipolar modes, labeled by three numbers $\{q,n,m\}$, whose associated complex
eigenfrequencies $\omega_{q,n,m}$ are the poles of the Mie coefficients
\cite{zambrana2015purcell}. In order to find the QNM resonances
in play in the previous configuration, we only look at the poles of the \emph{electric} Mie coefficients, because the dipole emitter is radially
oriented and therefore only couples to \emph{electric} modes (see
discussion at the end of Appendix \ref{App:B}). This consists in solving the
transcendental equation (see Eq.~(\ref{eq:app:mie_elec})):
\begin{equation}
(\varepsilon_s/\varepsilon_b) j_n(k_sa)\xi_n'(ka)=\psi_n'(k_sa) h_n(ka) \;,
\label{eq:mie_coeff}
\end{equation}
where all the functions and parameters are defined in Appendix~\ref{App:A}.
First note that Eq.~(\ref{eq:mie_coeff}) does not depend on $m$, which
means that multipolar modes with the same multipolar order $n$ but
different orbital number $m$ are
degenerate (\emph{i.e.} have the same eigenfrequency $\omega_n$). Therefore, the Lamb shift in Eq.~(\ref{eq:lambda_nf}) can be expressed
as a sum running on the multipolar order $n$,
\begin{equation}
\left. \frac{\Delta\omega}{\gamma_0} \right |_{\lambda_0} \simeq \sum_{n} -A_n\left(\frac{\lambda_n'}{\widetilde{\lambda}_0}\right)^2\frac{\lambda_n''}{2}\frac{\lambda_n'-\widetilde{\lambda}_0}{(\lambda_n'-\widetilde{\lambda}_0 )^2 + \lambda_n''^2} + B_n \;.
\label{eq:lambda_nf_bis}
\end{equation}
For each $n$, we find one solution $\omega_n$ of
Eq.~(\ref{eq:mie_coeff}) corresponding to the dominant pole, whose associated complex wavelength $\lambda_n$ is given in Table
\ref{tab:coeff} for $n=[1;5]$ (we still take the same Drude-Lorentz
model
for the permittivity of the silver nanosphere \cite{rakic1998optical}
as for the previous numerical simulations). The corresponding $A_n$ and $B_n$ terms are left
as free parameters and they are set by
fitting each multipole contribution $n$ of
Fig.~\ref{fig:multipolar_contribution} with the formula
$-A_n\left(\frac{\lambda_n'}{\widetilde{\lambda}_0}\right)^2\frac{\lambda_n''}{2}\frac{\lambda_n'-\widetilde{\lambda}_0}{(\lambda_n'-\widetilde{\lambda}_0)^2
  + \lambda_n''^2} + B_n\;,$ in Fig.~\ref{fig:QNM_contribution} (b)-(f). The discrepancy out of resonance that can be seen in Fig. \ref{fig:QNM_contribution}
(b)-(f) is due to the fact that
Eq.~(\ref{eq:omega_nf}) is valid only in the vicinity of the resonance frequencies
$\omega_\alpha$ and that we ignored the $\omega_0$ dependency of $B_\alpha$.
The values of the $A_n$ and $B_n$ parameters that result from the fit are given in Table \ref{tab:coeff} for $n = [1;5]$. Note
that the value of the amplitude $A_n$ decreases as $n$ increases, showing that the
resultant coupling between the emitter and the QNM resonance $n$ is less and less important.

In Fig.~\ref{fig:QNM_contribution} (a), we compare
the Lamb shift given by Eq.~(\ref{eq:lambda_nf_bis}) using the five QNM
resonances $n = [1;5]$ with the previous total Lamb shift calculated by computing
Eq.~(\ref{eq:Mie}) (black curve in Fig.~\ref{fig:multipolar_contribution}).
We can see that the
analytical formula Eq.~(\ref{eq:lambda_nf_bis}) based on the QNM resonances of the
plasmonic resonator qualitatively reproduces the Lamb
shift resonance when only a few dominant resonances are taken into account, but the
convergence could be further improved by increasing the number of QNM
resonances (see also \cite{PhysRevA.88.011803,Bakhti2014113}
where it is shown that a few set of QNM resonances is enough to reproduce
the scattering properties of a particle).
Moreover, this simple analytical formula clearly evidences that the Lamb
shift resonance results from the coupling of the quantum emitter to
the resonant modes of the nanoparticle. 

Finally, it is interesting to note that this resonant
coupling induces a \emph{positive} Lamb shift $\Delta \omega = \omega - \omega_0 >
0$ (around $340 \, \text{nm}$ in the configuration under study, see
Fig.~\ref{fig:multipolar_contribution} or Fig.~\ref{fig:QNM_contribution} (a)), which was first predicted in the case of silver \cite{PhysRevA.12.1448} and sodium
\cite{PhysRevA.32.2030} surfaces (see also \cite{klimov1996radiative} where a similar effect was reported
in the case of a
dielectric microsphere). This positive Lamb shift leads to a \emph{repulsive van der Waals potential} as long as the atom
remains in its excited state, which was shown experimentally with \emph{excited} cesium atoms in the presence of a sapphire surface
\cite{PhysRevA.51.1553,PhysRevLett.83.5467}.

\begin{table}[htbp]
\centering
\caption{\bf QNM complex wavelengths and fitting parameters.}
\begin{tabular}{cccc}
\hline
$n$ & $\lambda_n \text{(nm)}$ & $A_n$ & $B_n$ \\
\hline
$1$ & $375.6 + 15.5\mathrm{i}$ & $95.7$ & $-7.6$ \\
$2$ & $358.2 + 14.0\mathrm{i}$ & $57.8$ & $-8.2$\\
$3$ & $353.7 + 14.1\mathrm{i}$ & $38.3$ & $-6.1$\\
$4$ & $351.7 + 14.2\mathrm{i}$ & $24.5$ & $-4.1$\\
$5$ & $350.5 + 14.2\mathrm{i}$ & $15.0$ & $-2.6$\\
\hline
\end{tabular}
  \label{tab:coeff}
\end{table}

\section{Predictions about the Lamb shift}
\label{predictions}
\subsection{Blue-shift of the resonance}

\begin{figure}[h!]
\centering
\includegraphics[width=\linewidth]{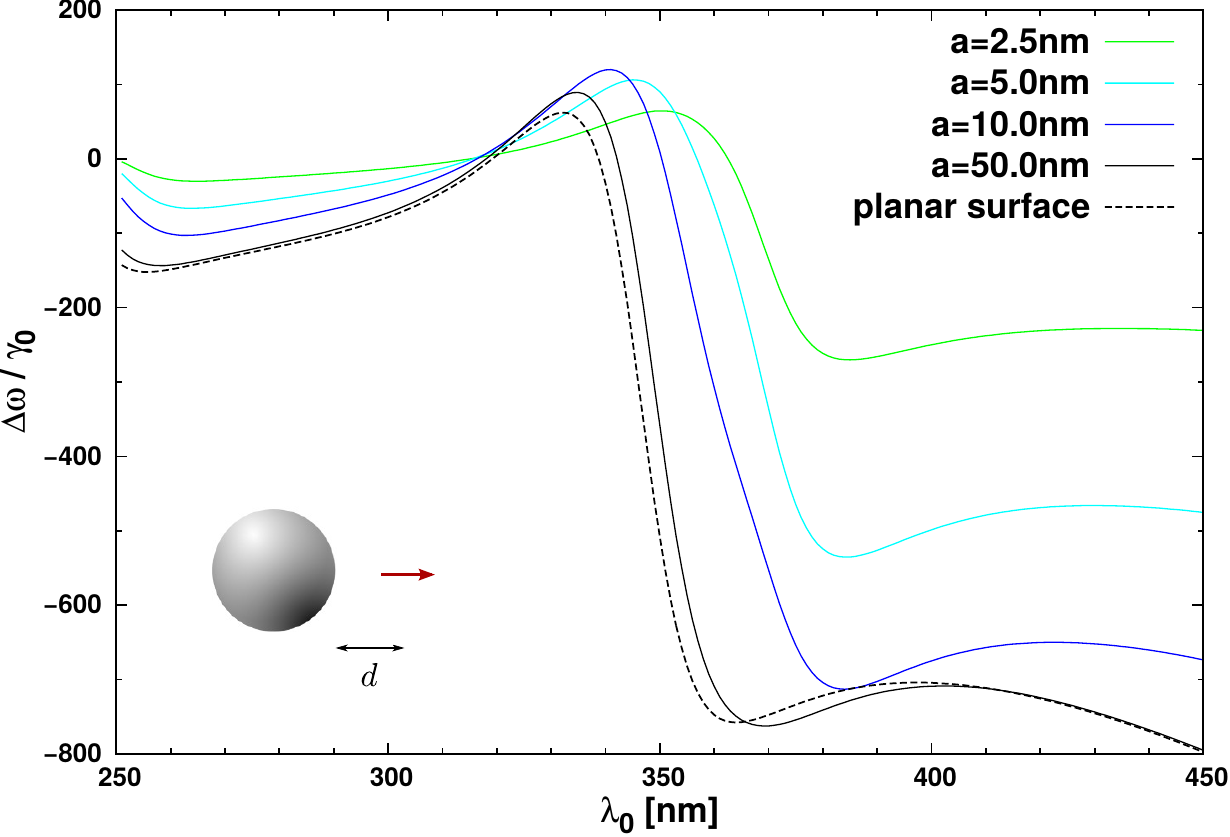}
\caption{Numerical simulations of the normalized Lamb shift $\Delta \omega/\gamma_0$ as a function of the transition
  wavelength $\lambda_0$ for a perfect electric dipole emitter with radial orientation and located at
  $d = 5\,\text{nm}$ from a silver nanosphere (red arrow), for different radii
  $a$ (full lines). The asymptotic case of a planar silver surface
  (Eq.~(\ref{eq:surface_freq})) is also plotted
  (dashed line). The refractive index of the
homogeneous background is $n_b=1$. A Drude-Lorentz
model for the silver permittivity is used according to
\cite{rakic1998optical}. The Lamb shift is computed by taking $n_{\text{cut}}=10$ except
  for the case $a=50\,\text{nm}$ where $n_{\text{cut}}=50$ in order to
  converge.}
\label{fig:radius}
\end{figure}

In this section, we show how the size of the nanoparticle affects the
position of the Lamb shift resonance. We still consider the case of a silver nanosphere.
We plot in Fig. \ref{fig:radius} the normalized Lamb shift as a
function of the transition wavelength for different particle radii
(full lines). The
asymptotic case of a planar surface is also
plotted (dashed line) according to the following expression \cite{PhysRevA.12.1448,PhysRevA.32.2030}:
\begin{equation}
\frac{\Delta\omega^\perp}{\gamma_0}=-\frac{3}{16k^3}\,\frac{|\varepsilon_s|^2-1}{|\varepsilon_s+1|^2}\,\frac{1}{d^3} \;,
\label{eq:surface_freq}
\end{equation}
which is valid in the non-retarded regime and for
an emitter oriented perpendicular to the surface. In this
case, the dipole emitter couples to the surface plasmon mode which comes from the
infinite density of states of the high order modes (around $\lambda_0
\simeq 340\,\text{nm}$ for a planar silver surface).

In sharp contrast with a nanosphere characterized by a purely dipolar response, we predict
in the near-field of the nanosphere a \emph{blue}-shift of the Lamb shift resonance as 
the radius of the nanosphere increases (see Fig. \ref{fig:radius}).
To understand this feature, let us recall that as the radius increases, each
plasmon resonance is
red-shifted and the dipole emitter couples to higher-order
multipoles \cite{ColasdesFrancs:08}. The displacement (blue-shift) of the Lamb shift resonance then results
from the \emph{interference} between these different modes. Therefore, this
effect will only exist if the dipole emitter is located in the near-field of
the nanoparticle, so that it will be able to excite \emph{several}
modes and to get this interference effect, resulting then in a blue-shift of the resonance. 

Thus, it can be observed in Fig. \ref{fig:radius} that in the near-field of the nanoparticle, a precise
engineering of this resonant coupling between the quantum emitter
and the plasmon resonances is possible. For instance, the
transition wavelength at which the Lamb shift is suppressed is
$\lambda_0=363 \, \text{nm}> \lambda_0=357\, \text{nm} > \lambda_0=350\, \text{nm} >
\lambda_0=342\, \text{nm} > \lambda_0=339 \, \text{nm}$ for the
radii $a=2.5 \, \text{nm}$, $a=5 \, \text{nm}$, $a=10
\, \text{nm}$, $a=50 \, \text{nm}$ and the case of the planar silver
surface respectively.
The tuning of this interaction is of current interest
\cite{aljunid2016atomic,chan2016tuning}, and we suggest
that thanks to their highly tunable optical properties, metallic
nanoparticles can also be used to tune and shape the Lamb shift of a
nearby quantum emitter through a control of their geometry, but also spatial organization and
environment, which can all be investigated through Eq.~(\ref{eq:num_sim}).

\subsection{Gold dimer nanoantenna}

In order to make a realistic calculation of the Lamb shift, let
us now consider a gold dimer with a dipole emitter located at the
center of the nanogap. This configuration is now experimentally
realizable using DNA templates \cite{busson2012accelerated,bidault2016picosecond}. 
To compute the Lamb shift, we take the parameters corresponding to \cite{punj2015self}: the nanoparticles radius is
$40\,\text{nm}$, the nanogap is $6\,\text{nm}$, and the effective refractive index surrounding the nanoparticles is
$n_{\text{eff}} = 1.5$; the fluorescent
molecule is an \emph{Alexa Fluor 647} dye, which presents an emission
peak around $\lambda_0 = 670\,\text{nm}$ with $40\,\text{nm}$ width; its total decay rate in the
homogeneous solution is measured at $\gamma_0 = 2.63 \,
\text{ns}^{-1}$ \cite{regmi2015nanoscale}.

\begin{figure}[htbp]
\centering
\includegraphics[width=\linewidth]{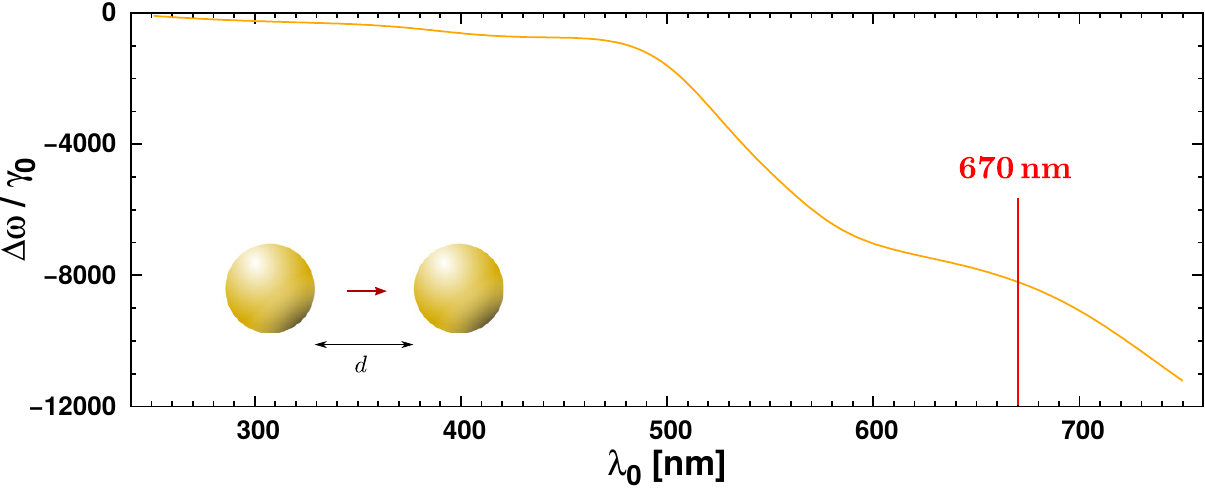}
\caption{Numerical simulations of the normalized Lamb shift $\Delta\omega/\gamma_0$ as a function of the transition
  wavelength $\lambda_0$ for a perfect electric dipole emitter with parallel
  orientation and located in the center of a gold dimer antenna
of radius $40\,\text{nm}$ and $6\,\text{nm}$ gap (red arrow). The refractive index of the
homogeneous background is $n_b=1.5$. A Drude-Lorentz
model for the gold permittivity is used according to
\cite{rakic1998optical}. The Lamb shift is
computed by taking $n_{\text{cut}}=40$.}
\label{fig:dimer}
\end{figure}

The Lamb shift spectrum of such a configuration with a dipole emitter
of parallel orientation is shown in
Fig. \ref{fig:dimer}. At $\lambda_0 = 670\, \text{nm}$, the normalized
Lamb shift computed with Eq.~(\ref{eq:num_sim}) is $\Delta\omega/\gamma_0=-8200$, which is \emph{outside} of
the range of the radiative linewidth, and therefore suitable for 
direct observation (the numerical simulations --- not shown here --- give
a radiative decay rate enhancement $\gamma_{r}/\gamma_0=1700$ at $\lambda_0 = 670\,
\text{nm}$). In order to find the Lamb shift of the dye, one
needs to multiply the value given by the numerical simulations by the reference \emph{quantum yield} $\phi_0=0.08$
in open solution (\emph{i.e.} without the antenna): $\Delta
\omega = \phi_0 \times (-8200) \times \gamma_0$.
The corresponding shift in terms of wavelength is given by the
following formula (valid if $\Delta \omega / \omega_0 \ll 1$):
$\Delta \lambda/\lambda_0 = - \Delta\omega/\omega_0$
where $\Delta\lambda = \lambda - \lambda_0$ with $\lambda$ the
new wavelength of the emitted photon. Thus, for the Alexa Fluor 647 dye, the relative shift is $\Delta\lambda /
\lambda_0 = 3.8\times10^{-3}$,
corresponding to a shift $\Delta\lambda = 2.5\,\text{nm}$.

Such a shift could be detected at room
temperature, by fitting the entire emission spectrum of
the molecule (see for instance \cite{ament2012single} where a shift of $\Delta\lambda \simeq
0.3 \,\text{nm}$ has been detected --- for the resonance spectrum of a
gold nanorod --- between neighboring Gaussian peaks
with width of about $50\,\text{nm}$ which is similar to our case here). One should 
also ensure that the spectral
dependence of the Lamb shift, decay rate enhancement and quantum
yield enhancement, do not vary appreciably in
the range used for fluorescence detection (the decay rate enhancement and
quantum yield enhancement spectra for the same configuration can be
found in \cite{punj2015self}, Fig.~3). 
In the future, it could be interesting to test the validity of the
weak-coupling approximation to quantify the Lamb shift in such a
configuration, by employing an other formalism suitable for
investigating the strong-coupling regime such as the one presented in \cite{Varguet:16}.

\section{Conclusion}

In this paper, we derived an exact multipole
formula, Eq.~(\ref{eq:num_sim}), to compute the Lamb shift induced by an
arbitrary set of resonant scatterers on a nearby quantum
emitter. In the case of a single silver nanoparticle, our numerical
simulations show that the dipole approximation
fails to account for the total Lamb shift spectrum in
the near-field region, and that one must include higher multipolar contributions. 
We furthermore adopted a Quasi-Normal Mode description of this
phenomenon, which provides a more physically intuitive understanding
of the induced Lamb shift as resulting from the coupling between the
quantum emitter and the resonances of the nanoparticle, and shows
that the total Lamb shift can be given by the sum of the independent
resonance contributions. These formulas also predict a displacement of the Lamb shift resonance
in the near-field to higher frequencies (blue-shift). Finally, a
calculation of the Lamb shift in a physically realistic configuration indicates that a direct detection may be possible for
fluorescent molecules embedded in a gold dimer nanogap.



\section{Appendix}
\subsection{Mie coefficients}
\setcounter{equation}{0}
\renewcommand{\theequation}{A{\arabic{equation}}}
\label{App:A}
In this Appendix, we give the expressions of the Mie coefficients in a
slightly different way then in \cite{bohren2008absorption} (where they
are called the scattering coefficients). 
By introducing $\varepsilon_s$ ($\mu_s$) and $\varepsilon_b$ ($\mu_b$) as the relative
permittivity (permeability) of the
sphere and the homogeneous background respectively, $k_s =
\sqrt{\varepsilon_s(\omega)}\omega/c$ and $k =
\sqrt{\varepsilon_b}\omega/c$, the Mie
coefficients of a sphere of radius $a$ take the form:
\begin{equation}
a_n = \frac{(\varepsilon_s/\varepsilon_b) j_n(k_sa)\psi_n'(ka)- \psi_n'(k_sa) j_n(ka)}{(\varepsilon_s/\varepsilon_b) j_n(k_sa)\xi_n'(ka)-\psi_n'(k_sa)h_n(ka)}
\label{eq:app:mie_elec}
\end{equation}
for the electric Mie coefficient of order $n$, and
\begin{equation}
b_n = \frac{(\mu_s/\mu_b) j_n(k_sa)\psi_n'(ka) -\psi_n'(k_sa)j_n(ka)}
{(\mu_s/\mu_b) j_n(k_sa)\xi_n'(ka)- \psi_n'(k_sa)h_n(ka)}
\end{equation}
for the magnetic Mie coefficient of order $n$, 
where $j_n(x)$ and $h_n(x)$ are respectively the spherical Bessel
functions and the first-type (outgoing) spherical Hankel functions, and $\psi_n(x)$ and $\xi_n(x)$ are the Ricatti-Bessel functions
defined as:
\begin{equation}
\psi_n(x) \equiv x j_n(x)
\end{equation}
\begin{equation}
\xi_n(x) \equiv x h_n(x) \;.
\end{equation}

\subsection{Analytical expressions of the dipolar and
  quadrupolar Lamb shift}

\setcounter{equation}{0}
\renewcommand{\theequation}{B{\arabic{equation}}} 

\label{App:B}
In this Appendix, we derive from Eq.~(\ref{eq:Mie}) analytical
expressions for the Lamb shift dipolar and quadrupolar contributions
for a sphere.
We consider the sphere placed in the $+z$ direction with respect to an electric dipole
emitter oriented either perpendicular to the surface of the sphere (orbital number $m = 0$,
dipole moment oriented on the $z$ axis) or parallel to the surface ($m =
1$, dipole moment oriented on the $x$ axis). Due to spherical
symmetry, the T-Matrix of the single sphere is a diagonal matrix $t$ composed of the Mie
coefficients of the sphere multiplied by $-1$. With a quadrupolar
assumption \cite{rolly2012promoting}:
\begin{equation}
t = -\text{Diag}(a_1,a_2,b_1,b_2) \;,
\end{equation}
with $a_1$ ($a_2$) the electric
dipolar (quadrupolar) Mie coefficient and $b_1$ ($b_2$) the
magnetic dipolar (quadrupolar) Mie coefficient defined in Appendix \ref{App:A},
\begin{equation}
f = [e_1,0,0,0]^t \;,
\end{equation}
with $e_1$ the incident electric dipole coefficient, and
\begin{equation} 
H^{(0,1)}= 
\begin{bmatrix}
A_{1,m,1,m} & A_{1,m,2,m} & B_{1,m,1,m} & B_{1,m,2,m} \\
A_{1,m,2,m} & A_{2,m,2,m} & B_{1,m,2,m} & B_{2,m,2,m} \\
B_{1,m,1,m} & B_{1,m,2,m} & A_{1,m,1,m} & A_{1,m,2,m} \\
B_{1,m,2,m} & B_{2,m,2,m} & A_{1,m,2,m} & A_{2,m,2,m} \\
\end{bmatrix} \;,
\end{equation}
where $A_{n,m,n',m'}$ ($B_{n,m,n',m'}$) the coupling coefficient from
the electric (magnetic) multipole order $n$ with orbital number $m$,
to the multipole order $n'$ with orbital number $m'$. Note that $H^{(1,0)}$ is the same as $H^{(0,1)}$ with all the $B$
coefficients multiplied by $-1$.
Employing the expressions of the coefficients $A$ and $B$ calculated
in \cite{rolly2012promoting} in Eq.~(\ref{eq:Mie}),
one gets for an electric dipole oriented perpendicular to the particle
surface ($m=0$):
\begin{equation}
\frac{\Delta\omega_1^\perp}{\gamma_0} = \frac{9}{2}\text{Im}\left[ a_1\frac{e^{2\text{i}k d}}{(kd)^6}(1-\text{i}kd)^2 \right]
\end{equation}
for the dipolar contribution and
\begin{equation}
\frac{\Delta\omega_2^\perp}{\gamma_0} = -\frac{9}{10}\text{Im}\left[ a_2\frac{e^{2\text{i}k d}}{(kd)^8}\left(-15\text{i}-15(kd)-25(kd)^2\right)^2 \right]
\end{equation}
for the quadrupolar contribution.
In the case of an electric dipole emitter oriented parallel to the
particle surface ($m=1$), the dipolar and quadrupolar contributions to the
Lamb shift read:
\begin{equation}
\begin{split}
\frac{\Delta\omega_1^\parallel}{\gamma_0} = & \frac{9}{8}\text{Im}\left[
  a_1\frac{e^{2\text{i}k d}}{(kd)^6}\left( 1-2\text{i}(kd) -
    3(kd)^2 + 2\text{i}(kd)^3+(kd)^4 \right) \right]\\
&-\frac{9}{8}\text{Im}\left[b_1\frac{e^{2\text{i}k
      d}}{(kd)^4}\left(\text{i} + (kd)\right)^2 \right]
\end{split}
\end{equation}
\begin{equation}
\begin{split}
\frac{\Delta\omega_2^\parallel}{\gamma_0} = & -\frac{15}{8}\text{Im}\left[ a_2\frac{e^{2\text{i}k
      d}}{(kd)^8}\left( 6\text{i} +
    6(kd)-3\text{i}(kd)^2-(kd)^3\right)^2 \right]\\
&+\frac{15}{8}\text{Im}\left[ b_2\frac{e^{2\text{i}k
      d}}{(kd)^6}\left(3-3\text{i}(kd)-(kd)^2 \right)^2\right]
\end{split}
\end{equation}
It is interesting to note in the case of a dipole emitter with
parallel orientation the presence of the magnetic Mie coefficients $b_1$ and $b_2$, which traduce
the cross-coupling between the electric dipole emitter and the
magnetic multipole resonances. This is not the case for a dipole
perpendicularly oriented whose multipolar Lamb shift contributions
only depends on the electric Mie coefficients, since the magnetic field produced by an electric
dipole is null along the dipole axis.

\section*{ACKNOWLEDGEMENTS}

The authors want to thank R\'emi Colom, Mahmoud Elsawy, Mauricio
Garcia-Vergara, Xavier Zambrana-Puyalto and
J\'er\^ome Wenger for fruitful discussions. E.~L. would like to thank the Doctoral School ''Physique et Sciences de la Mati\`ere'' (ED 352) for its
respective funding.

\bibliographystyle{apsrev4-1} 
\bibliography{sample}


\end{document}